# Mixture theory modeling for characterizing solute transport in breast tumor tissues

**Sreyashi Chakraborty[1], Alican Ozkan[2], Marissa Nichole Rylander,[2,3,4] Wendy A. Woodward[5], and Pavlos Vlachos[1]**

[1] Department of Mechanical Engineering, Purdue University, West Lafayette, IN 47907

[2] Department of Mechanical Engineering, The University of Texas at Austin, Austin, Texas 78712

[3] Department of Biomedical Engineering, The University of Texas at Austin, Austin, Texas 78712

[4] The Institute for Computational Engineering and Sciences, The University of Texas at Austin, Austin, Texas 78712

[5] Department of Radiation Oncology, MD Anderson Cancer Center, Houston, TX 77030

# Abstract

Delivery of nanoparticles to breast tumors involves a coupled analysis of the fluid and solute transport mechanisms in the microvasculature, across the vessel walls, and in the extravascular spaces. Tumor numerical models have been used to quantify solute transport with a single capillary embedded in an infinite tumor expanse, but measurements from different mammalian tumors suggest that a tissue containing a single capillary with an infinite intercapillary distance assumption is not physiological. The present study aims to investigate the limits of the intercapillary distance within which nanoparticle transport resembles solute extravasation in a breast tumor model as a function of the solute size, the intercapillary separation, and the flow direction in microvessels. Dextran transport properties obtained from an *in vitro* tumor platform containing a single vessel provided validation of the numerical model. Solute transport is modeled using mixture theory, applied to the nanoparticle accumulation and concentration decay in the tissue space for different vascular configurations. A comparison of a single capillary configuration (SBC) with two parallel cylindrical blood vessels (2 BC) and a lymph vessel parallel to a blood vessel (BC_LC) embedded in the tissue cylinder is performed for five solute molecular weights between 0.1 kDa and 70 kDa. The effects of counter flow (CN) versus co-current flow (CO) on the solute accumulation were also investigated and the scaling of solute accumulation-decay time and concentration was explored.

We found that the presence of a second capillary reduces the extravascular concentration compared to a single capillary and this reduction is enhanced by the presence of a lymph vessel. Co-current flow direction between two adjacent vessels led to nonhomogeneous nanoparticle distribution for larger particle sizes in the tissue space, while smaller particles (0.1 kDa and 3 kDa) showed the propensity to get trapped locally in the tissue during counter-current flow. Varying the intercapillary distance with respect to vessel diameter shows a deviation of 10-30 % concentration for 2 BC and 45-60% concentration for BC_LC configuration compared to the reference SBC configuration.

Finally, we introduce a non-dimensional time scale that captures the concertation as a function of the transport and geometric parameters. We find that the peak solute concentration in the tissue space occurs at a non-dimensional time, $T^*_{peak} = 0.027 \pm 0.018$, irrespective of the solute size, tissue architecture, and microvessel flow direction. This suggests that if indeed such a universal time scale holds, the knowledge of this time would allow estimation of the time window at which solute concentration in tissue peaks. Hence this can aid in the design of future therapeutic efficacy studies as an example for triggering drug release or laser excitation in the case of photothermal therapies.

**Keywords –** Solute transport, cancer tissue, mixture theory modeling

# Introduction

The total cost of cancer care in the United States is projected to increase by 39% from 2010 to 2020 (1). Primary areas of cancer research involve improving the efficacy of chemotherapeutic agents at the tumor sites and minimizing their toxic side effects in the non-target sites (2-4). Conventional chemotherapeutic agents (5) are non-specifically distributed in the body which limits the effectiveness of the drug dose and increases toxicity in normal cells. Drug carriers with hydrodynamic diameter 3-200 nm accumulate preferentially in tumors owing to the enhanced permeability and retention (EPR) effect (6) exploiting the wider pores in tumor vessels and the impaired lymphatic drainage in diseased tissues. The transport mechanism of these nanoparticles in tumors is a function of the hemodynamics, nanoparticle transport parameters (solute permeability, solute diffusivity, reflection coefficient) as well as the extravascular matrix properties (porosity, hydraulic conductivity). Before the binding/uptake by the cancer cell these particles overcome three major transport barriers: transport through microvasculature, translocation across the endothelial wall, and diffusion within the extracellular tissue matrix. Using a numerical model to investigate the nanoparticle transport mechanics could enable determination of the exact time interval between nanoparticle introduction and drug release to achieve desired therapeutic efficacy based on patient specific tumor measurements.

The majority of the existing multiscale models use the Darcy's law, Starling's law, and Poiseuille's law to analyze extravascular, trans-capillary, and intravascular transport respectively (7-12). Poiseuille's law cannot account for variations in capillary diameter and the inhomogeneous nature of blood. Deviations from Starling's law are expected when osmotic terms would include other endogenous solutes in addition to proteins. Darcy's law does not include the dependence of interstitial flow on local fluid chemical potential. Schuff et al. (13, 14) used mixture theory equations in an axisymmetric tissue geometry containing a concentric blood vessel and showed the dependence of extravascular fluid transport on chemical gradients in addition to hydrostatic pressure which was previously suggested (15, 16) and observed (17) but not commonly accounted for in previous transport models. In the present work, the mixture theory model is implemented in dual- tissue geometries to predict nanoparticle distribution in cancerous breast tissues over a wide range of particle sizes (0.5-15 nm) and molecular weights (0.1-70 kDa).

We hypothesize that nanoparticle distribution in breast tumors is a function of solute size, intercapillary separation, and flow direction and there exists a characteristic non-dimensional time, $T_{peak}^*$, for which solute concentration in the tissue space is maximum. We test this hypothesis by investigating the transport mechanisms of five solute types (0.1, 3, 10, 40 and 70 kDa) in tumor systems containing a single vessel (SBC) and compare with tumors possessing dual-vessel (blood capillaries only (2 BC), blood capillary and a lymph capillary (BC_LC)) tissue systems with varying intercapillary separation. The mixture theory equations are used for the first time and their predictive capability validated with measurements of dextran transport in an *in vitro* tumor platform containing multiple blood vessels

## Materials and Methods

The mixture theory equations model the transport (13, 14) of the fluid and solute in three distinct regions of a representative vascularized tumor geometry: a) in the intravascular space which consists of the plasma layer concentric with an inner core of red blood cells, b) across the capillary wall which is thin and semi-permeable and c) the extravascular space that comprises of the interstitial fluids and solutes flowing through a fibrous matrix. A finite element software package COMSOL 4.2 (COMSOL, Burlington, MA) was used to run the simulations. A schematic of the vascularized breast tumor configuration along with the transport pathways is shown in **Figure 1**. The blood vessel in **Figure 1** allows both intravasation and extravasation depicted by blue and red arrows respectively. The lymph vessel allows intravasation (blue arrow) only and drains the lymphatic fluid out of the tissue.

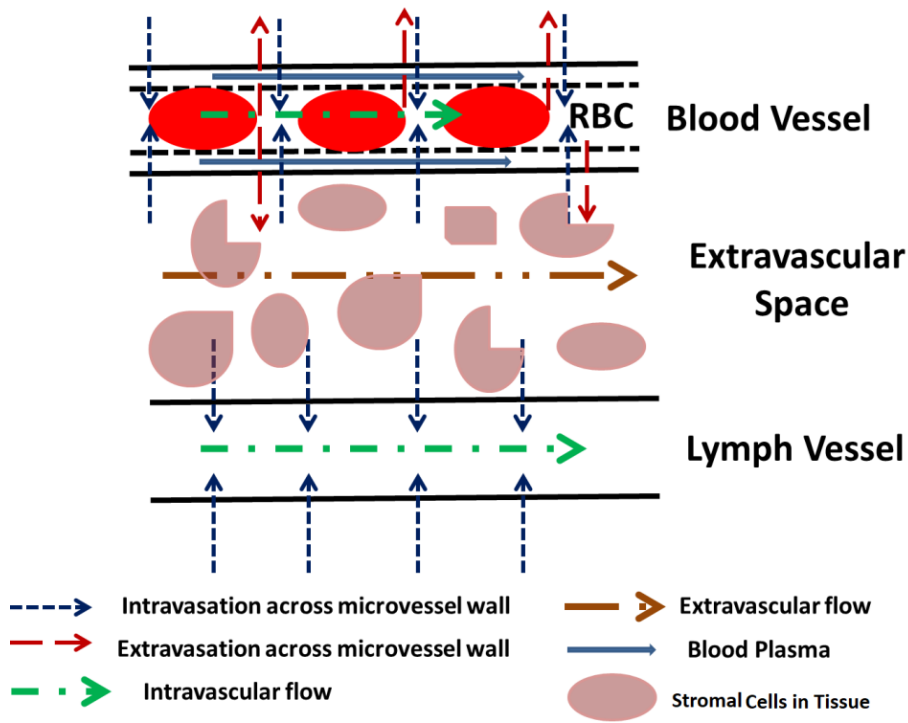

**Figure 1 : Schematic of a tissue containing blood capillary and lymph vessel. The arrows show the main transport mechanisms of a solute in the tissue. The extravascular space contains interstitial fluid flowing through a fibrous matrix. The blood vessel contains an inner core of red blood cells (RBC) surrounded by an outer plasma layer. The lymph vessel contains the interstitial fluid in it. No extravasation occurs in lymph vessels.**

## Mixture Theory Model Parameters

The mixture theory model requires fifteen input parameters that account for the mechanical properties of the porous matrix, type of the injected solute, and vascular geometry. For the current study, these parameters correspond to a human breast tumor and are presented in **Table 1**. The governing equations and boundary conditions for the mixture theory model have been derived in (13). A sensitivity analysis of the input parameters, calibration and subsequent validation of the model was carried out by Schuff et al. in (14). All the equations used in the simulations are explained in the Appendix A.

**Table 1: Input parameters for mixture theory model. Values are for different types of cancer tissues adopted from the literature. References are listed in the last column of the table.**

| Solute Dependent Parameters | | | | | | |
|---|---|---|---|---|---|---|
| Solute Molecular Weight (kDa), MW | 0.1 | 3.0 | 10.0 | 40.0 | 70.0 | |
| Hydrodynamic Diameter (nm) | 0.69 | 1.6 | 5.46 | 13.2 | 14.4 | (18-20) |
| Reflection coefficient, $\sigma$ | 0.00025 | 0.00025 | 0.02500 | 0.08600 | 0.14000 | (13, 14, 21, 22) |
| Solute Permeability coefficient , $P_d$ ($x10^{-8}$ m/s) | 800 | 174 | 70 | 33 | 30 | (13, 14, 23-25) |
| Diffusion Coefficient ($x10^{-11}$ m²/s), $D_f$ | 89.6 | 17.0 | 9.6 | 7.8 | 3.6 | (26-29) |
| Retardation factor, $R_F$ | 1.10 | 1.10 | 1.07 | 0.94 | 0.84 | (27-31) |
| Initial solute concentration (mol/m³), Co | 6.11 | 0.20 | 0.08 | 0.02 | 0.01 | (26) |
| Flow Parameters | | | | | | |
| Pressure gradient along blood vessel (Pa), dP | 2394 | | | | (32-35) | |
| Hydrostatic pressure in arteriole (Pa), Par | 4394 | | | | (32-35) | |
| Boundary tissue pressure (Pa), Po | 2700 | | | | (36-39) | |
| Osmotic Pressure gradient (Pa) | 2500 | | | | (13, 14, 18, 32) | |
| Hydraulic conductivity ($x10^{-15}$) (m²/Pa-s) | 400 | | | | (13, 14, 37, 40) | |
| Hydraulic permeability ($x10^{-10}$) (m/Pa-s) | 10 | | | | (13, 14, 37) | |
| Tissue porosity, $\emptyset$ | 0.4 | | | | (37, 41-43) | |
| Geometrical Parameters | | | | | | |
| Length of microvessels (mm), l | 1 | | | | (44, 45) | |
| Diameter of microvessels (µm), d | 10 | | | | (44, 46, 47) | |
| Diameter of tissue (µm), D | 200 | | | | (44, 46, 48-50) | |

## Experimental Validation of Mixture Theory Model

The accuracy of the computational model was confirmed with experimental measurements performed in a physiologically representative 3D vascularized *in vitro* tumor microenvironment. Essential model parameters such as tissue porosity, vessel porosity, solute permeability, and solute diffusivity were measured using the *in vitro* platform and implemented in the model (**Table 2**). The concentration-time histories were obtained from the mixture theory equations using the minimum, maximum, and mean values of the tissue parameters measured from *in vitro* platform. The simulation results were also compared with experimental measurements of from dextran transport in the same vascularized *in vitro* platform (**Figure 2**). Details explaining the fabrication and measurement processes can be found in the appendix B.

### Experimental and numerical comparison of concentration-time histories

Numerical simulations for the model validation studies were separately processed with identical tissue properties and boundary conditions as obtained from the experiment. The intensity-time histories were spatially averaged at a radial location of 600 $\mu$m. These were normalized by the maximum intensity inside the vessel at that time instant. For each of 3kDa and 70kDa solutes, transport was studied in N=3 tissue samples with identical fabrication parameters. The normalized intensity profile of dextran particles from these experiments corresponds to the normalized concentration from the numerical simulations (**Figure 2**). The error bars correspond to the experimental variability observed across 3 samples at each time instant. For 3kDa, the smaller solute, the experimental data till 1 hr matches well with the simulation curve from maximum values of input parameters. The deviation of experimental results from mean simulated values decreases with increasing time from 1hr to 2 hrs. The experimental data for the 70 kDa solute almost coincide with the mean simulation curve and is closely enveloped by the maximum and minimum simulation curves. To our knowledge, this is the first *in vitro* model that measured different porosity values in the extravascular and intravascular spaces.

**Table 2 : Parameters from the fabricated tissue platform used in the equivalent simulation**

| Parameters from fabricated tissue platform | Mean | Min | Max |
|---|---|---|---|
| Vessel diameter ($\mu$m) | 715 | - | - |
| Tissue diameter ($\mu$m) | 3000 | - | - |
| Tissue Porosity | 0.53 | 0.49 | 0.59 |
| Vascular Porosity | 0.4 | 0.37 | 0.43 |
| Solute Diffusivity ($m^2$/s) | 3 kDa: 25e-11<br>70 kDa: 4.3e-11 | 3 kDa: 20e-11<br>70 kDa: 3.7e-11 | 3 kDa: 30e-11<br>70 kDa: 4.9e-11 |
| Solute Permeability (m/s) | 3 kDa: 32e-8<br>70 kDa: 9e-8 | 3 kDa: 24e-8<br>70 kDa: 7e-8 | 3 kDa: 43e-8<br>70 kDa: 11e-8 |
| Hydraulic Permeability ($m^2$) | 1e-12 | - | - |

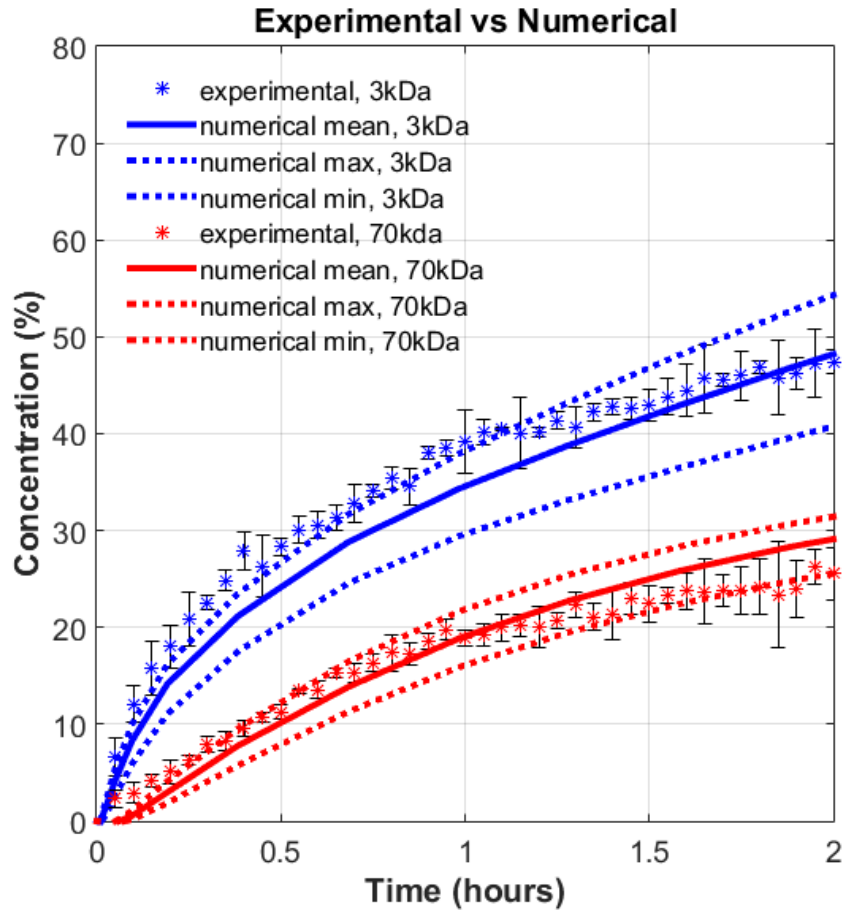

**Figure 2 : Comparison of experimental and numerical normalized concentration-time histories for 3kDa and 70 kDa solutes in a single capillary cancer tissue. a) The numerical curve under predicts the mean experimental concentration in earlier times and overpredicts in later times. b) The percentage deviation of the simulation concentration from the mean experimental concentration. For 70kDa the deviation is least at 1 hour while for 3kDa it is least at 2 hours.**

## Vascular Tissue Configurations and Test Matrix

In the current study three geometrical configurations of the vessels in the tissue are considered: the single blood capillary (SBC) configuration, the double blood capillary (2 BC) configuration, the blood capillary and the lymph vessel (BC_LC) configuration. These are shown in **Figure 3** along with the transport pathways. The tasks were split into 3 tests listed in **Table 3**. The flow direction is same (CO) in parallel microvessels for tests 1 and 3. The intercapillary distance (L) in the dual-microchannel configurations is 100 $\mu$m in tests 1 and 2 (27, 46, 50). The choice of flow parameters, tissue matrix properties and solute dependent parameters in each test are depicted in **Table 1**.

**Table 3 : Test matrix developed for conducting the study**

| Configuration | Flow Direction Type | Intercapillary separation ($\mu m$) | Solute Molecular weight (kDa) |
|---|---|---|---|
| **Test 1 : Effect of solute size** | | | |
| SBC | N/A | N/A | 0.1, 3.0, 10.0, 40.0, 70.0 |
| 2 BC | CO | 100 | 0.1, 3.0, 10.0, 40.0, 70.0 |
| BC_LC | CO | 100 | 0.1, 3.0, 10.0, 40.0, 70.0 |
| **Test 2 : Effect of flow direction in microvessels** | | | |
| 2 BC | CO | 100 | 0.1, 3.0, 10.0, 40.0, 70.0 |
| 2 BC | CN | 100 | 0.1, 3.0, 10.0, 40.0, 70.0 |
| BC_LC | CO | 100 | 0.1, 3.0, 10.0, 40.0, 70.0 |
| BC_LC | CN | 100 | 0.1, 3.0, 10.0, 40.0, 70.0 |
| **Test 3 : Effect of intercapillary separation** | | | |
| SBC | N/A | N/A | 3.0, 10.0 |
| 2 BC | CO | 10,50,100,250,1250 | 3.0, 10.0 |
| BC_LC | CO | 10,50,100,250,1250 | 3.0, 10.0 |

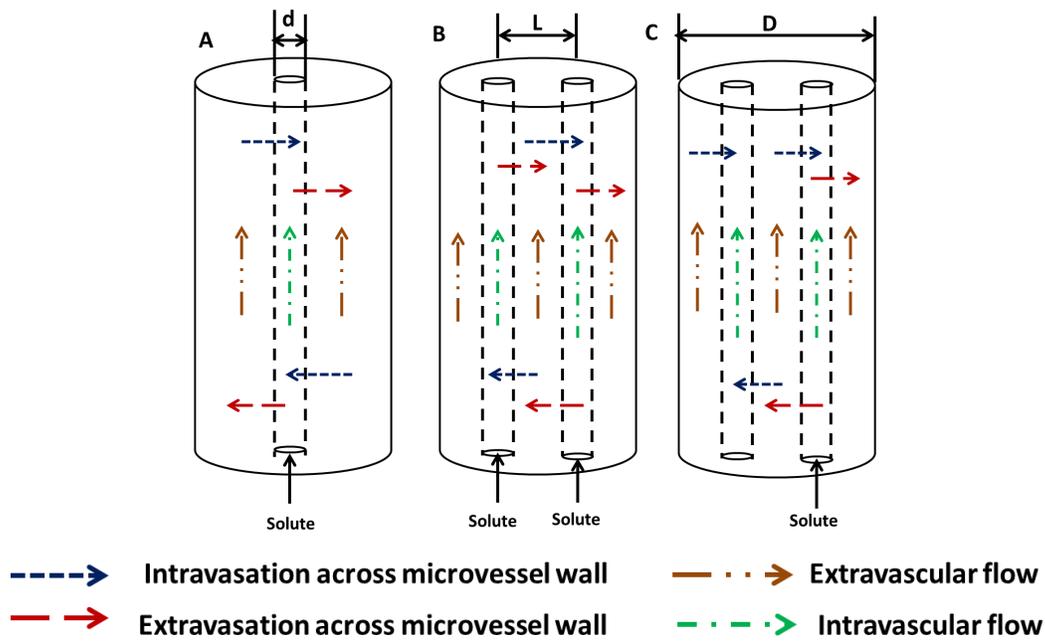

**Figure 3 : Three schematic tissue configurations with transport pathways that have been numerically modelled. A) single capillary embedded in tissue cylinder B) parallel flow blood capillaries in tissue cylinder C) lymph vessel parallel to blood vessel enclosed in tissue cylinder. d is the capillary diameter, L is the intercapillary separation and D is tissue diameter.**

**Table 4: Intercapillary distance from invivo tissues with capillary diameer, d=10 $\mu m$**

| Tissue Type | Intercapillary Separation (L, $\mu m$) | L/d | References |
|---|---|---|---|
| Rat mammary tumors | 50 | 5 | (44) |
| Rabbit neoplastic tissue | 101 | 10.1 | (48) |
| Mammary carcinoma | 80-135 | 8-13.5 | (49),(51),(50),(46) |
| Normal breast tissue | 215 | 21.5 | (51) |
| Human large intestine | 107 | 10.7 | (52) |
| Human colorectal tumor periphery | 54 | 5.4 | (53) |
| Human colorectal tumor center | 177 | 17.7 | (53) |

**Test 1: Effect of solute size on transport**

Five solutes (0.1 kDa, 3 kDa, 10 kDa, 40 kDa, 70 kDa) are injected at the inlet of the blood capillaries in each configuration to delineate the effect of solute size on transport. The geometry, hydraulic parameters and transport properties of the solute molecules in a breast tumor are in Table 1. The transport properties of 0.1 kDa solute are estimated from the calibration model developed in (14).

**Test 2: Effect of flow direction in capillaries on transport**

The second test investigates the influence of co-current flow and counter-current flow in both 2BC and BC_LC tissue configurations and compares the solute transport mechanisms with the SBC configuration.

**Test 3: Effect of intercapillary separation on transport**

In the third test, the capillary separation (L) is varied with respect to blood capillary diameter (d), (L/d=1, 5, 10, 25, 125), to test the solute accumulation in the tissue space as compared to the SBC configuration. Due to longer computational time, this test is run for two solutes, 3kDa (representative of a therapeutic drug) and 10 kDa (representative of the size of drug carrying nanoparticle). The intercapillary separation, L, is defined as the shortest distance measured between all non-adjacent capillary pairs in a loop (44). The tissue diameter (D) is twice the separation value (D=2L) to maintain the same volume of tissue around each microvessel with respect to the SBC configuration. The average extravascular concentration is calculated at a radial distance 0.2L from the blood vessel wall in SBC configuration ($C_{SBC}$) as well as in the 2 BC configuration ($C_{2BC}$). The percentage concentration deviation shown in **Figure 6**a is compared across different values of intercapillary separation. The intercapillary distances for different tissue types in humans and small animals are recorded in **Table 4**. The blood capillary diameter, d, is 10 $\mu m$ in accordance with the values reported in literature across humans, mice and rats (44, 46, 47). The L/d for tissue types shown in **Table 4** lies between 5 and 21.5. So the analysis was performed for L/d=1, 5, 10, 25 and 125 for two solute molecular weights of 3.0 kDa and 10 kDa. For all the tests, the average extravascular concentration in the tissue volume surrounding the blood capillary is measured for each configuration. They are normalized by the maximum intravascular concentration in the blood capillary volume and the concentration-time history for each test is analyzed in the Results section.

## Non-Dimensionalization of the Convection-Diffusion-Decay process

The final objective of this paper is to assimilate results of tests 1,2 and 3 to provide a unique non-dimensional time at which tissue solute concentration is maximum. The solute concentration-time histories in specific radial locations of the tissue space are influenced by the drainage of the solutes in addition to the advection diffusion and decay processes as modelled by **Eq 1**. S is the surface area density of the micro vessels whose walls act as a sink in the tissue volume. The product of S and solute permeability ($P_d^s$) is the decay constant k. We define $\tau$ as the time when the inlet concentration drops to 36.7% of the maximum inlet concentration. The first form of **Eq 1** is non-dimensionalized to find the time scales of the other transport mechanisms involved. Defining the following scaled variables $C_s* = \dfrac{c^s}{Co}; t* = \dfrac{t}{\tau}; x* = \dfrac{x}{L}; U_f* = \dfrac{u_f}{Uo}$, where $Uo$ is the average velocity of fluid in tissue. Substituting these in **Eq 1**; we get the non-dimensional form as shown in **Eq 2**.

$$\phi \frac{\partial c^s}{\partial t} + R_F^s u_f . \nabla c^s + \nabla.(-\phi D_{tissue} \nabla c^s) = -kc^s$$

**Eq 1**

$$\phi \frac{\partial c^s}{\partial t} + R_F^s u_f . \nabla c^s + \nabla.(-\phi D_{tissue} \nabla c^s) = -P_d^s S c^s$$

$$\frac{\phi}{\tau} \frac{\partial C_s*}{\partial t*} + R_F^s \frac{Uo}{L} U_f* . \nabla C_s* + \frac{D_{tissue}}{L^2} \nabla.(-\phi \nabla C_s*) = -kC_s*$$

**Eq 2**

The effect of $\tau$ on the solute accumulation in a tissue space is modified by the number of microvessels (n) present around it and the intercapillary separation (L/d) between them. The solute dependent timescales, obtained from Eq 2 are the modified input time scale ($n\tau \frac{L}{d}$), the diffusion timescale ($\frac{L^2}{D_{tissue}}$) and the decay timescale ($\frac{1}{k}$). These values for each solute in a SBC configuration are shown in **Table 5**. Since different timescales are dominant at different phases and radial locations of transport for differing solute types, a sum of all the solute dependent time scales is used to non-dimensionalize the time of solute accumulation and decay in tissue as shown in Eq 3.

$$T* = \frac{t}{n\tau \dfrac{L}{d} + \dfrac{L^2}{D_{tissue}} + \dfrac{1}{k}}$$

**Eq 3**

The extravascular concentration is rescaled to account for the variable solute molecular weight (Mw), solute density ($\rho$), tissue porosity $\emptyset$ and varying intercapillary separation (L/d) as defined by **Eq 4**.

$$C* = \frac{\rho}{\phi Mw} \cdot \frac{L}{d}$$

**Eq 4**

**Table 5 : Solute dependent timescales that influence concentration in tissue across time**

| Solute (kDa) | Input timescale ($\tau$) (s) | Diffusion timescale ($\dfrac{L^2}{D_{tissue}}$) (s) | Decay timescale (1/k) (s) |
|---|---|---|---|
| **0.1** | 284.5 | 25.4 | 125 |
| **3** | 440.1 | 133.4 | 574.7 |
| **10** | 604.8 | 241.7 | 1428.6 |
| **40** | 1302 | 328.2 | 3030.3 |
| **70** | 2319 | 778.9 | 3333.3 |

# Results

## Test 1: Effect of solute size in three configurations for fixed intercapillary separation

The effect of solute size in the double blood capillary (2 BC) and the blood capillary -lymph vessel (BC_LC) configurations compared to the single capillary (SBC) is shown in **Figure 4**. The tissue volume surrounding each capillary is equal. The concentration-time history essentially shows an initial accumulation period until the solute reaches its maximum concentration in the tissue space, then it is followed by a concentration decay. It is observed in the accumulation phase of the SBC configuration (**Figure 4**a inset) the 0.1 kDa, 3 kDa and 10 kDa solutes attain peak concentration with 41.3%, 40.3% and 30.7% of their maximum intravascular concentrations at 0.04, 0.07, and 0.12 hour respectively whereas heavier solutes like 40 kDa and 70 kDa attain 26.9% and 30.5% of their maximum intravascular concentrations at 0.22 and 0.35 hour respectively.

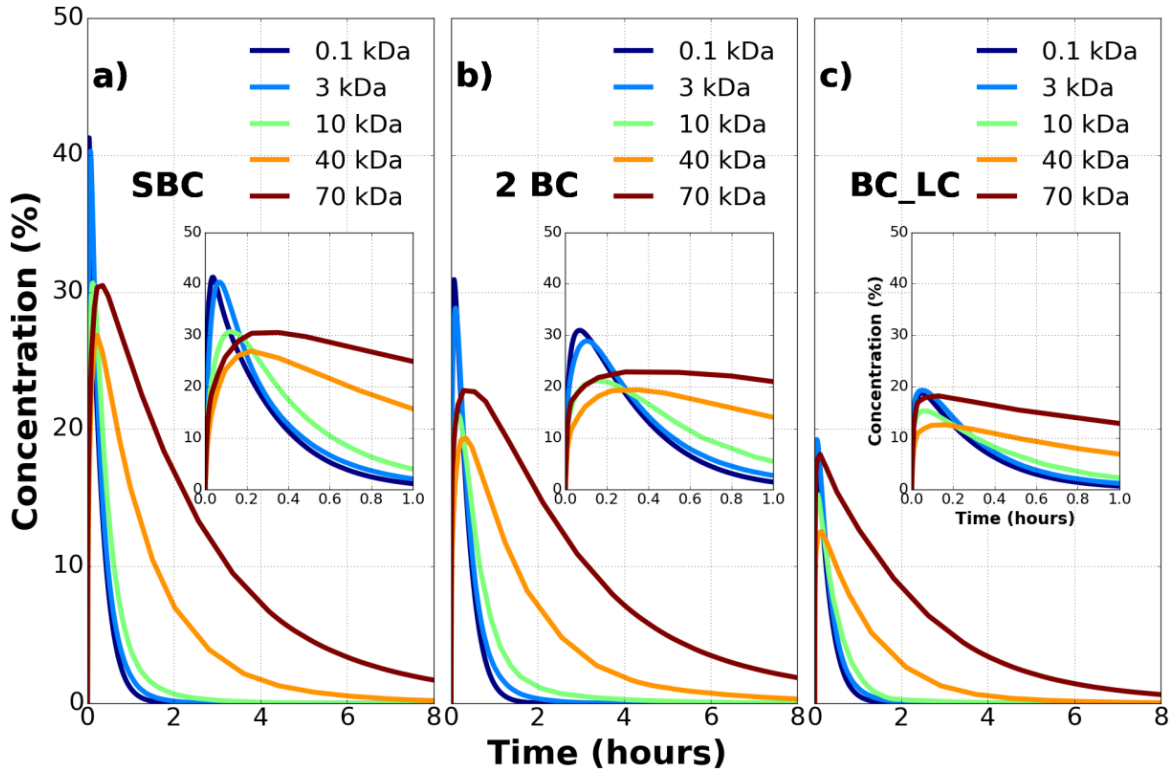

**Figure 4:Extravascular concentration-time history of five solutes in a) the SBC configuration b) The 2 BC configuration and c) the BC_LC configuration. Solute concentration in the tissue space decreases with time and varies with solute size. The inset in each subplot is a magnified view to show the concentration variation at earlier times for all three vessel configurations. Extravascular solute concentration reduces in a double blood capillary (2BC) configuration and decreases further in a lymph and blood capillary (BC_LC) configuration compared to single capillary (SBC) configuration.**

The time taken for 0.1 kDa, 3 kDa and 10 kDa solutes to reduce to 10% of their maximum intravascular concentrations are 0.43, 0.47 and 0.62 hour respectively whereas the same for the 40kDa and 70kDa solutes are 1.57 and 3.25 hours respectively. The peak tissue concentration for 0.1 kDa, 3 kDa, 10 kDa, 40 kDa and 70 kDa solutes decreases by 25%, 28%, 31%, 28% and 25% respectively in the 2 BC configuration (**Figure 4**b inset) and by 55%, 52%, 50%, 53% and 40% respectively in the BC_LC configuration (**Figure 4**c inset) with respect to the SBC configuration. In comparison to the SBC tissue peak concentration, the peak occurs at later times (**Figure 4**b) in 2 BC configurations (3 kDa: 0.11 hour vs 0.07 hour; 10 kDa: 0.16 hour vs 0.12 hour; 40 kDa: 0.35 hour vs 0.22 hour) and at earlier times (**Figure 4**c) in BC_LC configurations (3 kDa: 0.05 hour vs 0.07 hour; 10 kDa: 0.07 hour vs 0.12 hour; 40 kDa: 0.16 hour vs 0.22 hour). For the smallest solute 0.1 kDa, the concentration attains peak value later compared to its SBC counterpart (0.04 hour) in both the 2 BC (0.07 hour) and BC_LC (0.05 hours) tissue spaces. On the contrary the largest 70 kDa solute attains peak concentration earlier in both 2 BC (0.29 hour) and BC_LC (0.13 hour) extravascular spaces compared to its SBC counterpart (0.35 hour). The 70 kDa solute, however, exhibits a faster onset of concentration decay both in the 2 BC and BC_LC configuration (SBC: 0.35 hour; 2 BC: 0.29 hour; BC_LC: 0.13 hour) while a delayed concentration decay is seen for the 0.1 kDa solute (SBC: 0.04 hour; 2 BC: 0.07 hour; BC_LC: 0.05 hour).

## Test 2: Effect of flow direction in microvessels

The microvessel flows considered in test1 are in the same axial direction and are called co-current (CO) flows. They are compared with oppositely directed axial flows in the microvessels which are called counter current (CN) flows.

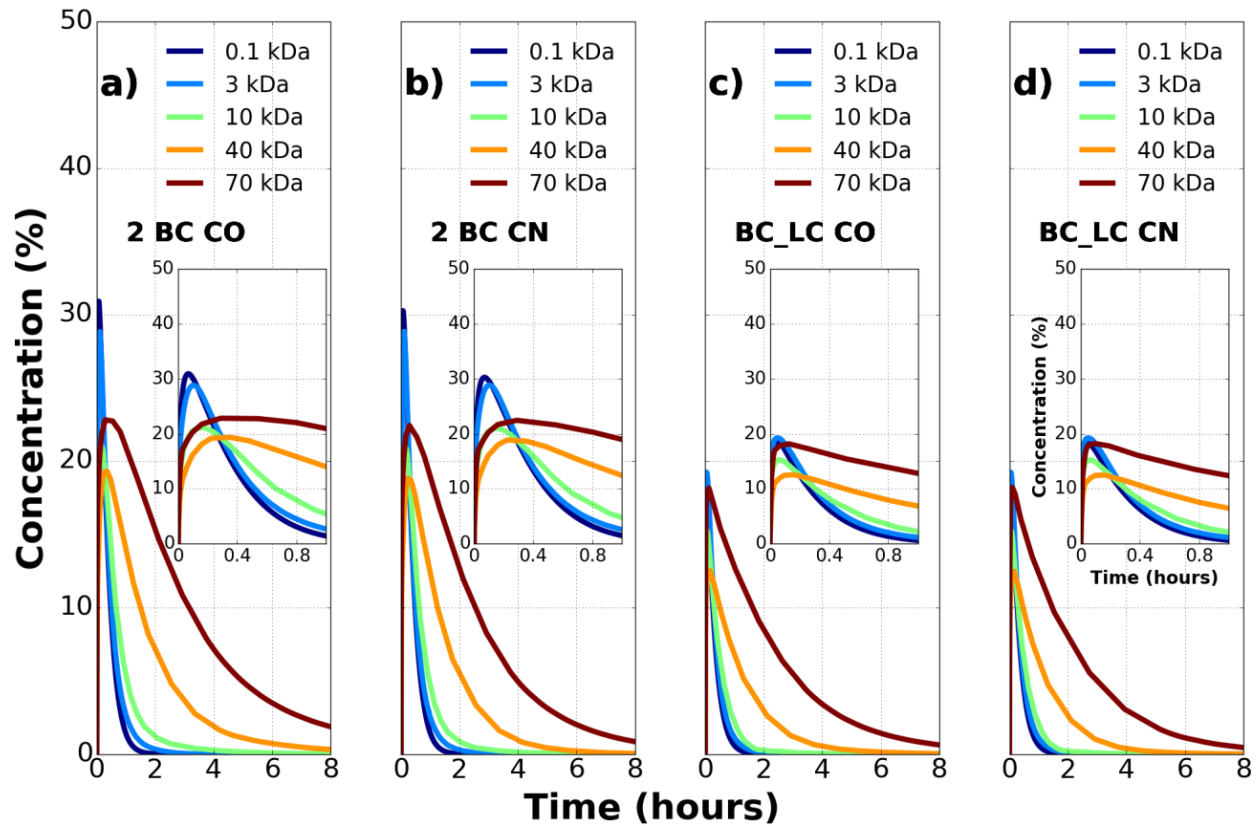

**Figure 5:** Extravascular concentration-time history of five solutes for a) Co-current (CO) flow in microvessels in 2 BC configuration b) Counterflow (CN) in microvessels in 2 BC configuration c) Co-current (CO) flow in microvessels in BC_LC configuration and d) Counterflow (CN) in microvessels in BC_LC configuration Counterflow (CN) reduces the solute concentration in tissue space more than co-current flow (CO) in parallel capillary configuration. The effect is more pronounced for the 2BC configuration at earlier times.

**Figure 5** compares the CN flow with the CO flow for 2 BC configuration and BC_LC configuration respectively. In both configurations there is no difference between the two flow types during the solute accumulation phase in the tissue. As the tissue concentration decays, the extravascular solute concentration is less in CN flow at later times compared to CO flow. The percentage reduction in concentration is more pronounced for larger (10 kDa: 0.5%; 40 kDa: 1.7%; 70 kDa: 2.5 % ) solutes in 2 BC configuration (**Figure 5**a, **Figure 5**b). A similar observation is made in **Figure 5**a, **Figure 5**b for heavier (10 kDa: 0.12%; 40 kDa: 0.4%; 70 kDa: 0.5%) solutes but the difference is less in BC_LC configuration (**Figure 5**c, **Figure 5**d) compared to the 2 BC configuration.

### Test 3: Effect of intercapillary separation on transport of 3 kDa and 10 kDa solutes

The intercapillary separation (L) was varied with respect to the vessel diameter (d) in the next set of tests for the 2 BC and BC_LC configurations. We calculated the average extravascular concentration at a radial distance 0.2L from the blood vessel wall in SBC configuration ($C_{SBC}$) as well as in the 2 BC configuration ($C_{2BC}$) where L is the intercapillary separation (**Figure 6**a). The percentage deviation between these two terms is plotted in **Figure 6**b across different values of non-dimensional intercapillary separation (L/d=1, 5, 10, 25, 125) for 3 kDa and 10 kDa solute. The dotted lines represent L/d values of 5 (yellow) and 21.5 (purple) which are the lower and upper limits of normalized intercapillary separation in breast tumors. L/d values above 21.5 are typically found in normal (non-diseased) tissues.

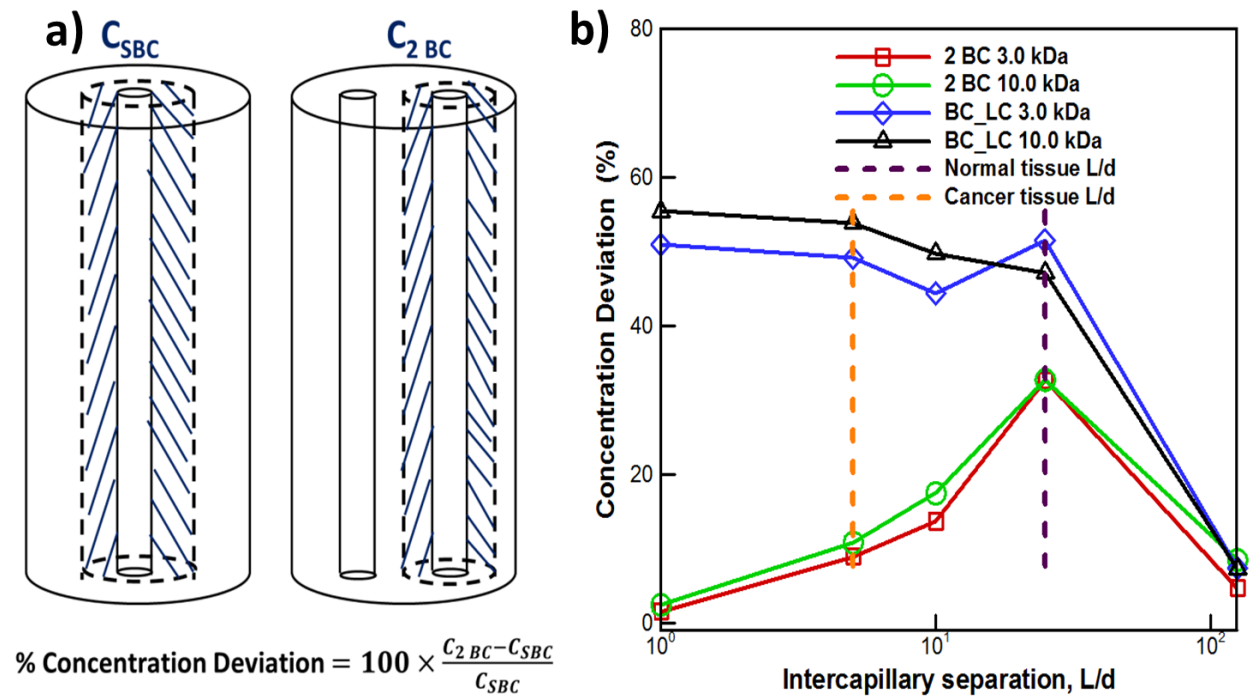

**Figure 6 : a) Schematic showing the setup and the calculation of % concentration deviation. b) Intercapillary separation (L/d) between 5 and 25 shows significant deviation of solute concentration in 2 BC and BC_LC configurations compared to SBC configuration for 3 kDa and 10 kDa solute sizes.**

## Double Blood Capillary (2 BC) embedded in tissue cylinder

The 2 BC configuration shows minimum deviation (3 kDa: 1.5%; 10 kDa: 2.5%) from the SBC configuration for L/d=1 (**Figure 6**b). This is because the spacing between two blood capillary walls is so small that they effectively function as a single capillary with twice the original capillary diameter. So the solute accumulation almost the resembles that in a SBC configuration. The solute concentration deviation is substantial (9-33 %) for L/d=5-25 which is the range of interest as depicted in **Table 4**. The isolated capillary assumption will not hold true for extravascular solute accumulation in this regime. The deviation (3 kDa: 4.7%; 10 kDa: 8.5%) reduces for L/d=125 because the large spacing between capillary walls minimizes the cumulative effect of the two blood vessels on the peak tissue concentration.

## Blood Capillary and Lymph Capillary (BC_LC) embedded in tissue cylinder

The BC_LC configuration (**Figure 6**b) shows minimum deviation (3 kDa: 7.4%; 10 kDa: 7.3%) from the SBC configuration for L/d=125 due to the same reason as the 2 BC configuration. But with decreasing L/d the sink action of the lymph vessel become increasingly dominant resulting in 44%-55% deviation of the maximum solute concentration in tissue volume from that in the corresponding SBC configuration in the L/d regime of 1 to 25.

## Non-dimensional Time vs Peak Non-dimensional Concentration Analysis

The results discussed in the previous cases have shown that the variation of solute size, microvessel arrangement, number of microvessels and intercapillary separation all contribute differently to the solute accumulation time vs solute decay time in the tissue space. Hence the non-dimensional extravascular solute concentration and non-dimensional time defined in **Eq 4** and **Eq 3** respectively were calculated for all test cases to account for variations of these four parameters. The non-dimensional profiles for fixed L/d=1, 5, 10, 25, 125 were plotted in **Figure 7**a. All concentration peaks lie within T*=0.1 shown by the dotted black line. So, in **Figure 7**b the scaled concentration time-histories from T*=0 to T*=0.1 were analyzed. The peaks were extracted and plotted in **Figure 7**c. The red curves that correspond to the largest intercapillary distance (L/d=125) have $T^*_{peak}$ values which are one order of magnitude less than the average $T^*_{peak}$. This is because for a large L, the second vessel does not contribute to the solute accumulation in the measurement location which is at a distance 0.2L from the first blood vessel. The concentration gradients across each capillary wall dynamically change the extravascular flux across the wall and for a large L the solute may get trapped within a certain distance of the capillary. The non-dimensional equations do not account for these and hence the deviation of $T^*_{peak}$ for L/d=125. It was concluded that the average non-dimensional time at which the peak concentration occurs in all configurations for all solutes is $T^*_{peak}$=0.027 ±0.018 (**Figure 7**c).

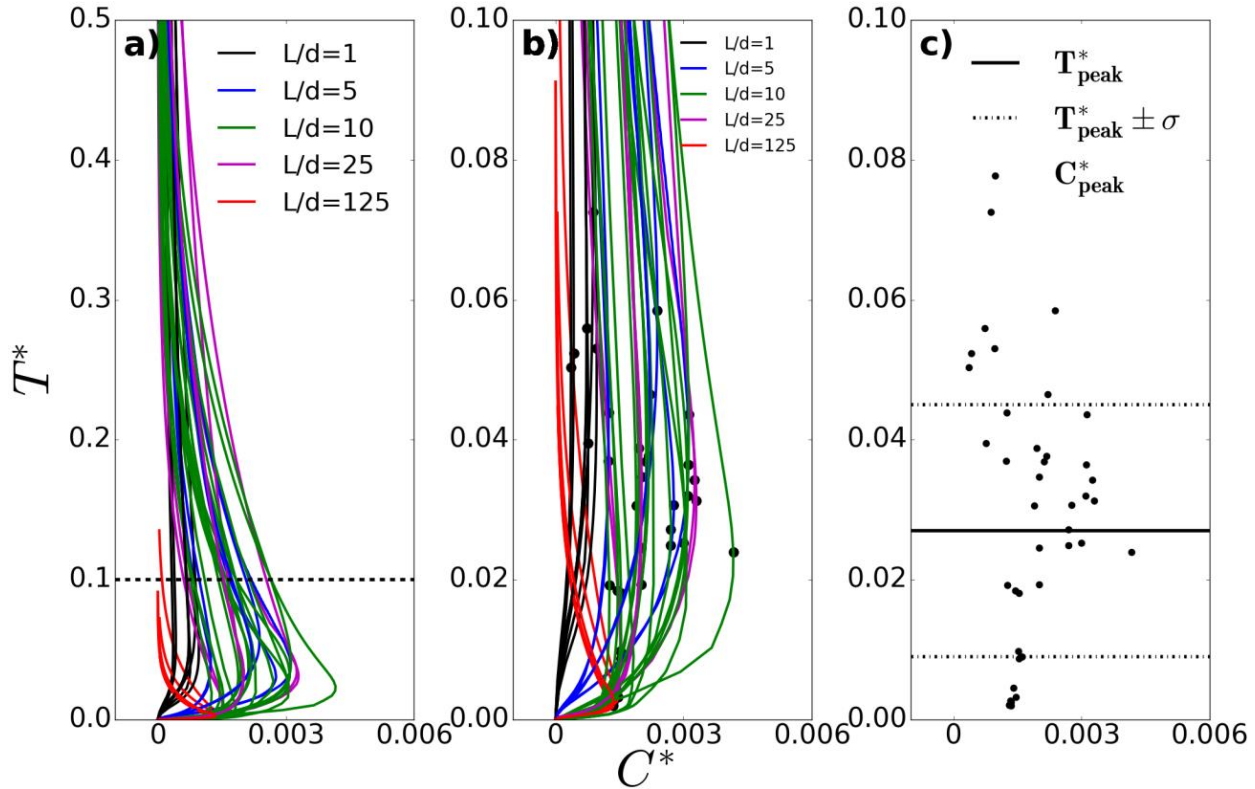

**Figure 7 :** **a) Non-dimensional concentration time history (T\* vs C\*) for all solute sizes in all tissue configurations with varying intercapillary separation b) A magnified view of the non-dimensional concentration time history to identify time occurrences of peak concentrations (C\*) c) The peak concentration (C\*) values are plotted separately to find an average non-dimensional time of occurrence as 0.027 ($T^*_{peak}$)±0.018 ($\sigma_{std}$).**

## Discussion

We present a numerical model for breast tumor that can predict passive transport of nanoparticles across a multilayer barrier when the tissue architecture and nanoparticle properties are specified. The solute size (molecular weight and hydrodynamic diameter) influences its extravascular concentration dynamically across time. Smallest solutes are cleared faster from the tissue but they are also susceptible to getting trapped in the recirculation zone set up by countercurrent blood flow in adjacent vessels (54, 55). High clearance rates measured for solutes <10 kDa during *in vivo* measurements in tumors validate this observation. The therapeutic outcome of breast cancer drugs (Doxorubicin: 0.54 kDa; Cisplatin: 0.3 kDa) having similar molecular weight as the smallest solute investigated here can now be predicted for patient-specific tumor biopsies.

Heavier nanoparticles (50-200 kDa) are preferred vehicles for the tumor location specific targeting and drug delivery(55, 56). According to the results, heavier solutes with hydrodynamic diameter (5-14 nm) take longer to attain maximum accumulation at a specific tissue location and are also removed slowly ((57, 58)). Presence of adjacent blood vessels with counter-current flow accelerate the clearance process owing to drainage from both ends of the tissue. Presence of a lymph vessel reduces the magnitude of their peak concentration considerably owing to high permeability cross the lymph capillary wall. In various *in vitro/in vivo* studies the lymph wall is shown to allow unidirectional flow only (59-62) that facilitates better drainage. Easy removal of all sizes of drugs through the lymphatic system contributes to role of lymphatics in worse response(63).

Blood vessels have been mostly modelled as non-porous structures permeable to solutes embedded in the flow (64, 65). To our knowledge, this is the first numerical model that account for endothelial porosity directly measured from dextran transport in an *in vitro* breast tumor model. Additionally, this work shows inclusion of multiple vessels in a tumor numerical model is necessary to accurately measure transport phenomena. The SBC assumption works only for tissues where capillaries are so close (L/d=1) that they act as a single vessel, e.g when nearby lymph vessels collapse (59, 60, 62, 63) during metastasis of some cancers or if they are so far apart (L/d=125) that the solute flux from one does not reach the other.

The dextran transport investigated using a fabricated 3D microfluidic platform measured tissue porosity, endothelial porosity, nanoparticle permeability and nanoparticle diffusivity. Simulations driven by these parameters showed a close correspondence of numerical and experimental concentration-time histories. These parameters, when reported in literature, span over several orders of magnitude. The wide range can be attributed to the complex *in vivo* measurements whose intrusive nature would perturb the tissue microenvironment (17, 66). Thus, the ability to measure these parameters *in vitro* can be used to design future non-invasive transport investigation studies.

Condensing all the effects of tissue architecture, solute and fluid transport properties, there exists a unique time $T_{peak}^*$ at which nanoparticle concentration in the tissue is maximum. Previous analytical solutions of a convective-dispersive solute transport equation with time-dependent inlet boundary condition (67, 68) have shown the dependence of time constant on the input timescale, convection timescale, decay timescale and the diffusion timescales but did not account for varying intercapillary separation. Later Chapman et al. and other researchers (8, 12, 69) modeled transport in tumors characterized by intercapillary separation but did not account for the dynamic change of hydraulic permeability as a function of concentration gradients which in turn modulates the extravascular solute flux. The presented work, for the first time, analyzes the solute concentration in the tissue in the light of mixture theory equations for varying solute types, two parallel microvessels, differing flow directions in parallel vessels and tissue architecture and proposes a non-dimensional time at which solute concentration is maximum in the tissue.

Since this approach non-dimensionalizes the intercapillary separation (L) with the vessel diameter (d), $T_{peak}^*$ can be predicted for tissues ranging over several scales and also for different disease stages (cancer vs normal). This prediction would aid in efficient endothelial targeting, triggering drug release and laser excitation for photothermal therapies (70-72). The $T_{peak}^*$ estimation can hugely impact the clinical landscape as it would customize treatment based on tumor specificity. Future studies with varying nanoparticle design, changing dosage, presence of magnetic targeting, receptor binding can all be implemented first to the mixture theory model whose predictions would increase the efficacy of the targeted drug delivery in patient specific tumors.

The major limitation associated with the study is simplification of the complex vascular network. The tortuosity and diameter variation of the microvessels were neglected. The extracellular matrix was considered stationary and not allowed to deform. A zero flux boundary condition was prescribed at each microvessel outlet which deviates from the physiological condition where a constant solute flux is drained to other organs like the liver from the microvessel outlets.

## Conclusion

The study described in this paper focuses on quantification of solute transport across parallel blood vessels and initial lymph vessels in the light of mixture theory. Transport of nanoparticles to the targeted tumor volume is defined by the transport through microchannels, diffusion across endothelium and transport within the porous matrix, all of which were accounted for in the presented work. The results show that the solute size strongly influences its own rate of removal and rate of accumulation in the tissue. The flow physics in the extravascular space facilitate tissue drainage of nanoparticles depending on the solute size, the intercapillary separation and the microvessel arrangement in the tissue.

A unique non-dimensional time $T^*_{peak}$ was reported for the first time. This is the time at which peak concentration of a nanoparticle occurs at any tissue location, irrespective of the solute size, the intercapillary separation and the microvessel flow direction. The knowledge of the nanoparticle introduction time, tissue mechanical properties and solute dependent properties will allow, in future, to design *in vitro* tissue models testing varying nanoparticle designs and concomitantly, predict for patient specific tumors, the appropriate time of drug release that can substantially improve drug efficacy.

## Acknowledgement

We would like to acknowledge funding provided by the National Institutes of Health Grant 5R21EB019646. We would like to thank Professor Eric Nauman for introducing us to the mixture theory model for solute and fluid transport in tissues.

## Conflict of Interest

There is no conflict of interest to report.

# Appendix A: Mixture theory model equations

### Fluid transport Equations

The arterial pressure, Par, is specified at the inlet while the prescribed hydrostatic pressure difference, $\nabla P$ balances the viscous stresses to govern the fluid flow inside the capillaries according to **Eq. 1**.

$$-\nabla P + \mu \nabla^2 u_f = 0 \qquad\qquad \textbf{Eq. 1}$$

The fluid flux, $q_e$, across the capillary wall (of radius Ro) as shown in **Eq. 2** accounts for a) the hydrostatic pressure difference across the capillary wall due to the hydraulic conductivity, $\tilde{L}_p$, b) a constant osmotic pressure gradient, $\sigma^p(\tilde{\pi}_i^p - \tilde{\pi}_e^p)$, due to protein molecules; and c) a variable osmotic pressure that depends on the concentration difference of the injected solute (of molecular weight $M_w^s$ and density $\rho_T^s$) in the intravascular (i) and the extravascular (e) space of a fibrous matrix with porosity $\emptyset$.

$$q_e = \tilde{L}_p[(P_i\,|_{\,r=Ro} - P_e\,|_{\,r=Ro}) - \sigma^p(\tilde{\pi}_i^p - \tilde{\pi}_e^p) - (\overline{P} + \frac{A}{\emptyset})\frac{M_w^s}{\rho_T^s}\sigma^s(c_i^s\,|_{\,r=Ro} - c_e^s\,|_{\,r=Ro})] \qquad \textbf{Eq. 2}$$

The fluid transport in the extravascular space is influenced by a) the hydrostatic pressure difference in the tissue space, b) the hydraulic permeability of the tissue, $k$ and c) the solute concentration gradients, $\nabla c^s$ in the tissue space as depicted in **Eq. 3**. The retardation factor, $R_F^s$, is the ratio of the solute velocity and the fluid velocity in the tissue space. A constant pressure Po is applied to the tissue boundary.

$$-\nabla P - \frac{U_f}{k} + (\overline{P} + \frac{A}{\emptyset})\frac{M_w^s}{\rho_T^s}(1 - R_F^s)\nabla c^s = 0 \qquad\qquad \textbf{Eq. 3}$$

### Solute transport Equations

The initial solute concentration, $C_o$, is used to prescribe the concentration at the inflow according to **Eq. 4**. The concentration time history for the five solutes prescribed at the blood vessel inlet is shown in **Figure 8**. The concentration has been normalized by the peak intravascular concentration of each solute.

$$C_{inlet} = \left\{ \begin{array}{l} \dfrac{C_o \mathrm{t}}{15}, (t < 15s) \\[2ex] 0.5 C_o(e^{\frac{tA1}{60}} + e^{\frac{tA2}{60}}), (t > 15s) \end{array} \right\}$$

$$\qquad\qquad \textbf{Eq. 4}$$

$$A1 = -7.23(M_w^s)^{0.38}, A2 = -0.062\exp(-3.66e^{-5}M_w^s) - 0.0035\exp(-5.78e^{-7}M_w^s)$$

The advection-diffusion equation governs the solute transport inside the microvessels where $D_{sf}$ is the solute diffusion coefficient is shown below:

$$\frac{\partial c}{\partial t} + u.\nabla c + \nabla.(-D_{sf}\nabla c) = 0 \qquad \textbf{Eq. 5}$$

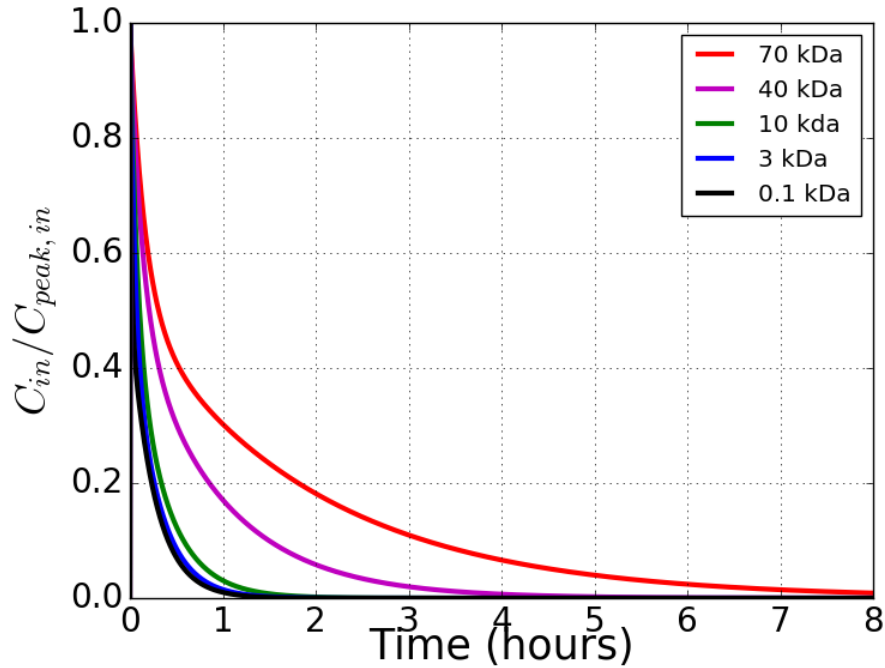

**Figure 8 : Concentration vs Time profile of five solutes at the blood vessel inlet.**

The solute particles carried by the fluid flux and those which permeate into the tissue space due to the concentration difference across the capillary wall constitute the solute flux ($J_s$) across it. **Eq. 6** is a modified version of Starling's law where $P_d^s$ is the solute permeability coefficient and $\sigma^s$ is the reflection coefficient.

$$J_s = P_d^s(c_i^s\mid_{r=Ro} - c_e^s\mid_{r=Ro}) + (1-\sigma^s)\overline{c^s}q_e \qquad \textbf{Eq. 6}$$

In the porous matrix of the extravascular space the solute undergoes both advection and diffusion as shown in the governing transport **Eq. 7**.

$$\phi\frac{\partial c^s}{\partial t} + R_F^s u_f.\nabla c^s + \nabla.(-\phi D_{tissue}\nabla c^s) = 0 \qquad \textbf{Eq. 7}$$

The mixture theory equations are appropriate for this study since it shows the dependence of tissue mechanical properties like hydraulic conductivity on chemical gradients that is not captured by traditional transport models (13).

**Difference of lymph vessel input parameters from that of blood vessel**

Solute concentration is zero at the lymph capillary inlet. The osmotic pressure gradient due to protein molecules is absent in the lymph vessel, $(\sigma^p(\tilde{\pi}_i^p - \tilde{\pi}_e^p) = 0)$. The solute permeability coefficient, $P_d^s$, across the lymph capillary wall is twice its value in blood capillary wall shown in **Table 1** to account for the free permeability of lymph vessels to macromolecules. The fluid flux and the solute flux equations are modified to allow intravasation only.

## Appendix B: Fabrication of microfluidic platform

Type I collagen was used as the extracellular matrix of the tumor with a single integrated endothelialized blood vessel. Excised tendons from rat tails were dissolved in a pH 2.0 HCl solution for 12h at 23˚C. The solution was centrifuged at 30000g for 45 minutes and sterilized using 10% (v/v) chloroform for 24h at 4˚C. The mold for the *in vitro* tumor microfluidic platform with the embedded single vessel was fabricated as described in previous work (73, 74). PDMS housing was fabricated using common soft-lithography methods. Polydimethylsiloxane (PDMS) and curing agent was mixed with 10:1 ratio and baked at 75˚C for 1 hour. Hardened PDMS housing and the glass cover slip was plasma treated for 18W for 30 seconds. Plasma treated surfaces were assembled to create a permanent bonding. The housing was treated with 1% (v/v) polyethyleneimine in $dH_2O$ for 10 min followed by 0.1% (v/v) glutaraldehyde in $dH_2O$ for 20 min and washed with $dH_2O$ twice. Collagen solution of 7 mg/ml was prepared by neutralizing stock solution with 1X DMEM, 10X DMEM, 1N NaOH, and mixing $1x10^6$/ml MDA-MB-231 breast cancer cells uniformly in collagen which was then placed in the housing. A 22G (711µm) needle was inserted into the mold and after polymerization and the needle removal  a cylindrical vascular channel was created within the collagen.  $2x10^6$ TIME cells were injected into the vascular channel and exposed to flow preconditioning protocols for 3 days to form a confluent, aligned endothelialized vessel. As a result, *in vitro* platform shown in Figure 9 was fabricated.

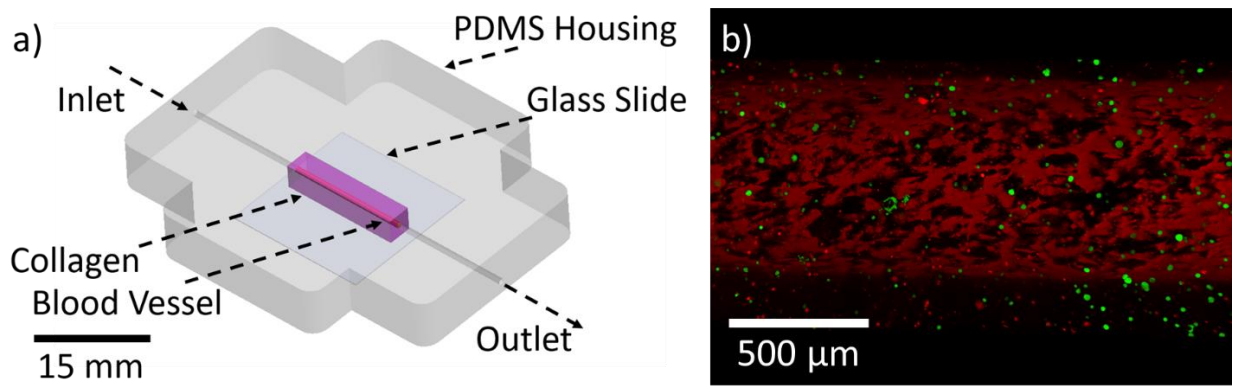

**Figure 9:** 3D vascularized *in vitro* microfluidic platform for experimental validation. a) CAD drawing of the platform. b) Confocal image of blood vessel with endothelial cells (red) surrounded by breast carcinoma cells (green).

Measuring mechanical properties from fabricated tissue platform

*Porosity* Vasculature porosity was measured using confocal microscopy and imaging mKate labeled TIME cells. Scanning electron microscopy (Zeiss, Super40) was used to image 3 fibrous matrix samples under 15kx, 20kx, and 25kx magnifications at three vertical planes. Tissue porosity was measured by applying a Frangi filter on obtained images.

*Solute Permeability* Obtained intensity profiles of 3 kDa and 70 kDa Dextran particles were used to calculate solute permeability coefficient as shown previously(75). 3kDa and 70kDa dextran particles were suspended in serum free endothelial basal medium at 10 µg/ml concentration and perfused in the vascular channel at 260 µL/min, which corresponds to 1 dyne/$cm^2$ physiological shear stress for tumor vasculatures at every 3 minutes for 2 hours. The transport of these solutes was imaged using a confocal microscope (Leica SP8, 10X magnification). Normalized intensity profiles from this images as a function of time were used to compare with normalized concentration profiles from the equivalent numerical simulations.

*Solute Diffusivity* Fluorescence recovery after photobleaching (FRAP) technique was used as described previously by Voigt et al. to measure the diffusion coefficient for a range of dextran molecular weights (4 kDa-150 kDa) for varying pH values, collagen concentrations and temperatures(76, 77). We selected diffusivity values of 3 kDa and 70 kDa for pH 7.6, collagen concentration 7mg/ml at $37^0$C from the database of the mentioned study.

*Hydraulic tissue permeability* Hydraulic permeability for collagen at a concentration of 7 mg/ml as is used in fabrication of the platform were collected from the existing literature on vascularized *in vitro* experiments (40, 78, 79).